\documentstyle[12pt]{article}
\def\g{\gamma}

\def\r{\rho}

\def\inf{\infty}
\def\x{\xi}

\def\bg{\begin{eqnarray}}
\def\ed{\end{eqnarray}}

\begin{document}
\begin{flushright}
SFU-Preprint-92-6 \\
\end{flushright}
\vskip.15in
\begin{center}
{\Large\bf Dynamical Self-mass for Massive Quarks}
\\ \vskip .70in
{\bf Zheng Huang and K.S. Viswanathan}\\
Department of Physics\\ Simon Fraser University\\
 Burnaby, B.C. \\ Canada V5A 1S6\\
\end{center}
\vskip .50in
\begin{abstract}
We examine dynamical mass generation in QCD with large current mass
quarks.  A renormalization group analysis is performed to separate
fermion self-mass into a dynamical and a kinematical part. It is shown
that the energy scale of the Schwinger-Dyson (SD) equation and the
effective gauge coupling are fixed by the current mass. The dynamical
self-mass satisfies a homogeneous SD equation which has a trivial
solution when the current mass exceeds  a critical value. We therefore
suggest that the quark condensate, as the function of the current mass,
observes
a local minimum around $e\Lambda _{QCD}$.
\end{abstract}
\vskip 1.0in
\today
\pagebreak
\section{Introduction}
Dynamical chiral symmetry breaking in QCD has been extensively studied
\cite{review}. The standard tool to study this problem is the
Schwinger-Dyson (SD) equation, i.\ e.\ the fermion gap equation. Since
one is usually interested in the chiral limit one studies the gap
equation in the limit of the vanishing current quark mass. In the
presence of a quark whose current mass is much larger than $\Lambda
_{QCD}$, the nature of solutions to SD equation may exhibit
non-analytic behavior in $m$. It has long been argued by Pagels
\cite{pagels} that if the current mass is large, it is possible that
$m\neq 0$ does not belong to an analytic extension of $m=0$. In this
article we explore this scenario using the renormalization group
equation (RGE) analysis.

Recently, Langfeld, Alkofer and Reinhardt \cite{lar} have reported on a
numerical study of the Schwinger-Dyson equation for massive quarks in
the background field of a classical vacuum solution. Besides the usual
spontaneous chiral symmetry breaking solution for $m=0$, they find
that the total quark condensate
as a function of the current mass exhibits a discontinuous drop at
$m_c\simeq 70 MeV$. This is interesting because not much has been
known about the behavior of quark condensates for quarks of large
current mass.

However, in a recent conference report \cite{zh} we have discussed the
problem of dynamical mass generation in QCD with large current mass
($m\gg \Lambda _{QCD}$). It was pointed out that using the
renormalization group analysis one can separate unambiguously the
quark self-mass (the part of the self-energy which commutes with $\g _5$)
 into a dynamical and a kinematical part. The
dynamical part $B_{D}$ satisfies a homogeneous SD equation but
with an effective coupling constant
$\bar{g}^2(t)\simeq (\ln \frac{m}{\Lambda_{QCD}})^{-1}$ and hence when
$m\gg \Lambda _{QCD}$ it describes weak coupling regime. As a
consequence, the SD equation for $B _D$ can be solved quite
reliably in a single effective gluon exchange and one finds that
$B _D$ has only a trivial solution when $m$ exceeds a critical
value $m_c$ which is of the order of
$2.7 \Lambda _{QCD}$. The kinematical part $B _K$, however,  is a
non-singular function of the current mass and satisfies an
inhomogeneous equation and hence is proportional to $m$. The total
fermion condensate can be written similarly as a sum of a dynamical
and kinematical part which is expected to take a drop near current
mass of the order of $m_c$\footnote{We have neglected the non-trivial
topological
gauge configurations.  Inclusion of them may lead to a contribution
from the gluon condensate. However, the essential feature of the behavior of
the total quark condensate should not be substantially affected by this.}.

The aim of this paper is to elaborate on our method which utilizes the
renormalization group techniques. The dynamical self-mass may be understood
as arising from an effective interaction with the gluons.  It would be of
interest to explore if the dip in condensate has any observable consequences
\cite{hvw}.
\section{The Renormalized SD Equation}
The quark self-mass $B(p^2)$ satisfies the following SD equation (the
gap equation) in the momentum space
\bg
B(p^2)=m+{{ig^2C_2(N)} \over 4}Tr\int {{{d^4k} \over {(2\pi )^4}}D^{\mu \nu
}(p-k)\gamma _\mu S(k)\Gamma _\nu (p,k)}
\label{1.1}
\ed
where $S(k)$, $D^{\mu\nu}(k)$ and $\Gamma _{\nu}(p,k)$ are the
complete quark, gluon propagators and the proper vertex respectively.
$m$ is the current mass; and $C_2(N)=\frac{N^2-1}{2N}$ for $SU(N)$.
The quark propagator $S$ is related to $B$ by
\bg
S^{-1}(k)=\not \!\! k+\not \!\! E(k)-B(k^2)
\label{1.2}
\ed
where $\not \!\! E(k)$ is the wave function correction factor.  Before
trying to solve (\ref{1.1}), we must rewrite it in terms of
renormalized quantities and specify a renormalization prescription.
We take the point of view that the renormalization constants in the
renormalized SD equation may be defined as in perturbation theory. We
also require that after carrying out the renormalization prescription
we get a finite SD equation. We adopt the dimensional regularization
and mass independent renormalization scheme \cite{thooft}. We define
renormalized quantities as follows
\bg
S(p;g,m,\xi ;\epsilon ) & = & Z_F(g(\mu );\epsilon )S_R(p;g(\mu ),m(\mu ),\xi
(\mu );\mu );
 \nonumber \\
D^{\mu \nu }(p;g,m,\xi ;\epsilon ) & = & Z_A(g(\mu );\epsilon )D_R^{\mu
\nu }(p;g(\mu ),m(\mu ),\xi (\mu );\mu );
 \nonumber \\
\Gamma _\mu (p;g,m,\xi ;\epsilon ) & = & Z_F^{-1}Z_A^{-1/2}\Gamma _\mu
^R(p;g(\mu ),m(\mu ),\xi (\mu );\mu ) \label{1.3} \\
B(p;g,m,\xi ;\epsilon ) & = & Z_F^{-1}B_R(p;g(\mu ),m(\mu ),\xi (\mu
);\mu ) \nonumber
\ed
where
\bg
g(\epsilon ) & = & Z_g(g(\mu );\epsilon )g(\mu );\nonumber \\
m(\epsilon ) & = & Z_m(g(\mu );\epsilon )m(\mu ); \nonumber \\
\xi (\epsilon ) & = & Z_\xi (g(\mu );\epsilon )\xi (\mu ).\nonumber
\ed
The renormalized quantities are functions of $g(\mu )$ and
$\epsilon$. $\x$ in (\ref{1.3}) is the gauge parameter. Substituting
(\ref{1.3}) into (\ref{1.1}) we obtain the renormalized integral
equation (in Euclidean $4-\epsilon$ dimensions)
\bg
B^R(p;\mu ) & = & Z_F(\epsilon )[Z_m(\epsilon )m(\mu )+Z_g^2(\epsilon
)Z_A^{1/2}(\epsilon )g^2(\mu )C_2(N)
 \label{4}\\
 & & \cdot {{\mu ^\epsilon } \over {4-\epsilon }}Tr\int {{{d^{4-\epsilon }k}
\over {(2\pi )^{4-\epsilon }}}\gamma ^\mu D_R^{\mu \nu }(p-k;\mu )S_R(k;\mu
)\Gamma _R^\nu (p,k;\mu )]}.
\nonumber
\ed
In (\ref{4}) we have written
$S_R(p;g(\mu ), m(\mu ), \x (\mu );\mu )$ as $S_R(p;\mu )$ for short.

Eq.\ (\ref{4}) is in general not self-consistent since its right hand
side contains divergent renormalization constants as
$\epsilon\rightarrow 0$. It has been suggested by Johnson, Baker and
Willey \cite{jbw} that a suitable choice of the gauge parameter $\x$
(Landau gauge) would lead to a finite $Z_F$ as
$\epsilon\rightarrow 0$. Subsequently, it would be possible to
require that the divergences arising from the integral cancel those
in $Z_m(\mu ,\epsilon )m(\mu )$ and make (\ref{4}) finite. Thus, a
consistent solution to (\ref{4}) should satisfy the condition of
finiteness of $Z_F(\mu ,\epsilon )$ as $\epsilon \rightarrow 0$. We
show below using RGE analysis that when the quark current mass is
large, this condition can be approximately satisfied.

\section{Renormalization Group Analysis}
Equation (\ref{4}) contains the exact propagators and vertex. If the current
mass is small, $B^R(p;g,m,\x ,\mu ,)$ may be expanded in powers of $m(\mu )$
\bg
B^R(p;g,m,\xi ;\mu ) & = & \underbrace {B_0^R(p;g,0,\xi ;\mu )}_{B_D^R}
\label{4.5}\\
& & +\underbrace {m(\mu )B_1^R(p;g,0,\xi ;\mu )+m(\mu )^2B_2^R(p;g,0,\xi
;\mu )+\cdots }_{B_K^R} \nonumber
\ed
where we call the first term $B_0$ as the `dynamical' part of $B^R$ and other
terms as the `kinematical' part. A non-vanishing dynamical part signals the
spontaneous chiral symmetry breaking. (\ref{4.5}) is just the chiral
perturbation
expansion, which can only make sense when the series is convergent or $m$ is
small ($m\ll \Lambda _{QCD}$). We would like to develop a scheme applicable
when $m>\Lambda _{QCD}$, when the power series solution does not converge. A
convenient method in this case is to apply RGE treatment. We may write
(\ref{4}) as a functional equation
\bg
F(p;g(\mu ),m(\mu ),\xi (\mu );\mu )=0. \label{5}
\ed
{}From (\ref{5}) it follows that
\bg
(\mu {\partial  \over {\partial \mu }}+\beta (g){\partial  \over {\partial
\kern 1ptg}}+\gamma _mm{\partial  \over {\partial \kern 1ptm}}+\delta (g)\xi
{\partial  \over {\partial \xi }})F=0.
\label{6}
\ed
In a mass-independent renormalization scheme, $\beta (g)$, $\gamma (g)$, and
$\delta (g)$ depend only on $g(\mu )$. Introducing $t$ by
\bg
t=\ln \frac{m_P}{\mu }
\label{7}
\ed
with $m_P=m(\mu =m_P)$ and observing that $F$ is homogeneous of order 1 in
$p$, $m$ and $\mu$, we have
\bg
({\partial  \over {\partial \kern 1ptt}}+p{\partial  \over {\partial \kern
1ptp}}+\mu {\partial  \over {\partial \mu }}-1)F(p;g,e^tm',\xi ;\mu )=0.
\label{8}
\ed
In (\ref{8}) $m'(\mu )=e^{-t}m(\mu )$. From (\ref{6}) and (\ref{8}), it
follows that $F$ satisfies the following RGE:
\bg
(-{\partial  \over {\partial \kern 1ptt}}-p{\partial  \over {\partial \kern
1ptp}} &+&\beta (g){\partial  \over {\partial \kern 1ptg}}+\gamma _mm'{\partial
 \over {\partial \kern 1ptm'}}\nonumber \\
  &+&\delta (g)\xi {\partial  \over {\partial \xi }}+1)F(p;g,e^tm',\xi ;\mu
)=0.
\label{9}
\ed
{}From (\ref{9}) it follows that
\bg
F(p;g,e^tm',\xi ;\mu )=e^tF(e^{-t}p;\bar g(t),\bar m'(t),\bar \xi (t);\mu )
\label{10}
\ed
where the effective parameters $\bar{g}(t)$, $\bar{m'}(t)$, and $\bar{\x}(t)$
are defined by
\bg
t=\int\limits_g^{\bar g(t)} {{{dx} \over {\beta (x)}}},
\label{11}
\ed
\bg
\bar m'(t)=m'(\mu )\exp \int\limits_g^{\bar g(t)} {dx{{\gamma _m(x)} \over
{\beta (x)}}},
\label{12}\ed
and
\bg
\bar \xi (t)=\xi (\mu )\exp \int\limits_g^{\bar g(t)} {dx{{\delta (x)} \over
{\beta (x)}}}.
\label{13}
\ed
It should be noted that even though $t$ is a function of $\mu$, the effective
coupling $\bar{g}(t)$ depends on $m_P$ only.
An explicit calculation of (\ref{11})-(\ref{13}) gives the following results
in one-loop approximation in the minimum subtraction scheme
\bg
\alpha (m_P)\equiv {{\bar g^2(t)} \over {4\pi }}={{2\pi } \over {\beta _0\ln
{{m_P} \over {\Lambda _{QCD}}}}},
\label{14}\ed
\bg
\bar m'(t)=e^{-t}m(\mu )\exp \int\limits_g^{\bar g(t)} {dx{{\gamma _m(x)} \over
{\beta (x)}}}=e^{-t}\bar m(t)=e^{-t}m(\mu =m_P)=\mu ,
\label{15}
\ed
and
\bg
\bar \xi (t)=1-{1 \over {\xi _0(\ln {{m_P} \over {\Lambda _{QCD}}})^{d_\xi }}},
\label{16}
\ed
where $\beta _0=\frac{33-2n_f}{3}$, $d_{\x}=\frac{39-4n_f}{2(33-2n_f)}$
\cite{narrison} and $\x _0$ is a constant of integration. It is interesting to
observe that $\bar{m'}(t)$ is just the scale parameter $\mu$. This arises
because we have defined $t$ by the on-shell current mass $m_P$.

The physical consequence of the RGE analysis follows from equation (\ref{10}).
Instead of solving the SD equation (\ref{1.1}) with a large current mass $m$
directly by approximation procedures for $\Gamma _\nu$ and $D^{\mu\nu }$, we
can solve, equivalently an effective SD equation govern by a much smaller mass
$\mu$ ($=\bar{m'}(t)$) and running coupling and gauge parameters. A suitable
approximation to these are given in (\ref{14})-(\ref{16}). It is seen from
(\ref{16}) that when $m_P\gg \Lambda _{QCD}$, $\bar{\x}(t)\simeq 1$ (Landau
gauge) and $\bar{g}(t)\rightarrow 0$ as $m_P\rightarrow \inf$.  Thus a large
current mass justifies the single gluon exchange approximation for the
effective SD equation and ensures the compatibility of the condition
$Z_F(\bar g,\bar \x , \epsilon )\rightarrow 1+O(\bar{g}^2)$ as
$\bar{\x}\rightarrow 1$,  whereas, it would have been hard to justify such an
approximation in (\ref{1.1}).

As a consequence of these results, the renormalized
Green's functions $D_R^{\mu\nu}$ and $\Gamma _R^\mu$ may be approximated by
\bg
D_R^{\mu \nu }(p;\bar g,\mu ,\bar \xi ;\mu ) & = & {{\delta ^{\mu \nu }-p^\mu
p^\nu /p^2} \over {p^2}}+O(\bar g^2);\nonumber \\
 \Gamma _R^\mu (p;\bar g,\mu ,\bar \xi ;\mu ) & = & \gamma ^\mu +O(\bar g^2).
\label{17}
\ed
Substituting (\ref{17}) into the effective SD equation as given in (\ref{10}),
we arrive at the RGE improved effective SD equation to  $O(\bar{g}^2)$
\bg
B(p^2)=Z_m(\bar g;\epsilon )\mu +3\bar g^2C_2(N)\int {{{d^{4-\epsilon }k} \over
{(2\pi )^4}}{{(2\pi \mu )^\epsilon } \over {(p-k)^2}}{{B(k^2)} \over
{k^2+B^2(k^2)}}}+O(\bar g^4),
\label{18}\ed
where we have used the notation
$B(p^2)\equiv B^R(p;\bar g, \mu,\bar{\x};\mu )$. Finally, from a solution to
(\ref{18}), we can reconstruct $B^R(p;g(\mu ),m(\mu ), \x (\mu );\mu )$
defined in Eq.\ (\ref{4}) from the following equation, which is easily derived
\bg
B^R(p;g(\mu ),e^tm(\mu ),\xi (\mu );\mu
)\quad\quad\quad\quad\quad\quad\quad\quad\quad\quad\quad\quad\quad \label{19}
\\
  =e^t\exp [-\int\limits_g^{\bar g(t)} {dx{{\gamma _F(x)} \over {\beta
(x)}}}]B^R(e^{-t}p;\bar g(t),\mu ,\bar \xi (t);\mu ).
\nonumber \ed

Now in (\ref{18}) the inhomogeneous term contains a mass of the order of
$\Lambda _{QCD}$, we can thus use a chiral perturbation theory to expand
$B^R(p;\bar{g}(t),m_0 ,\bar{\x}(t);\mu )$ as power series in $m_0$ (in our
case, of course, $m_0=\mu$. We use a different symbol to make this
decomposition
more obvious)
\bg
B^R(p;\bar g(t),m_0,\bar \xi (t);\mu ) & = & B_{dyn}^R(p;\bar g(t),0,\bar \xi
(t);\mu )\nonumber \\
 & &+m_0B_{kin}^R(p;\bar g(t),m_0,\bar \xi (t);\mu )
\label{20}\ed
where we have called the term independent of $m_0$ as the dynamical self-mass
and have lumped the rest as the kinematical part. In this way of splitting, it
is obvious that the kinematical part is a regular function of the $m_0$ and it
goes to zero as $m_0\rightarrow 0$. But what is crucial is that both $B_{dyn}$
and $B_{kin}$ depend on $m_P$ in a $non$-$analytical$ way through $\bar{g}(t)$.

Substituting (\ref{20}) into (\ref{19}) we obtain the following structure for
the self-mass for a heavy quark (taking $\gamma _F=0$)
\bg
B^R(p;g(\mu ),m(\mu ),\x (\mu );\mu) & = &
e^tB_{dyn}(e^{-t}p;\bar{g}(t))\nonumber \\
& & +m(\mu )\exp(-\int^{\bar{g}(t)}_{g}dx\frac{\gamma
_m(x)}{\beta (x)})B_{kin}(e^{-t}p;\bar{g}(t),m_0).  \label{21}
\ed
It is to be emphasized again that the separation of the self-mass into a
dynamical and a kinematical part is not the usual power series expansion in
current mass. In fact, substituting (\ref{20}) into (\ref{18}) we derive an
effective homogeneous SD equation for $B_{dyn}$
\bg
B_{dyn}(p;\bar g(t))=3\bar g^2(t)C_2(N)\int {{{d^4k} \over {(2\pi )^4}}{1 \over
{(p-k)^2}}{{B_{dyn}(k^2)} \over {k^2+B_{dyn}^2(k^2)}}}+O(\bar g^4)
\label{23}\ed
while $B_{kin}$ satisfies
\bg
B_{kin}(p;\bar g(t))=1+3\bar g^2C_2(N)\int {{{d^4k} \over {(2\pi )^4}}{1 \over
{(p-k)^2}}{{B_{kin}(k^2)} \over {(k^2+B_{dyn}^2(k^2))^2}}(k^2-3B_{dyn}^2)}.
\label{24}\ed
Both (\ref{23}) and (\ref{24}) have been studied in the context of chiral
symmetry breaking \cite{fukuda}.

\section{Solutions and Discussions}
The effective SD equations  for the heavy quark differ from the those for the
light quark where the chiral symmetry breaking is of concern.  In the
latter case the gauge coupling constant is a function of the momentum
transfer. In the infrared range the coupling becomes arbitrarily large and a
non-trivial solution exists for light quarks. In (\ref{23}) and (\ref{24}) the
coupling constant is fixed and allows us to seek for a solution in the wider
range (not just the asymptotic solution in the limit $p^2\rightarrow \inf$).
However, a non-trival solution to (\ref{23}) should not be referred to as the
signal of chiral phase transition.

Eq.\ (\ref{23}) is equivalent to the following differential equation
\bg
x^2{{d^2B_{dyn}(x)} \over {dx^2}}+2x{{dB_{dyn}(x)} \over {dx}}+\lambda
{{xB_{dyn}(x)} \over {x+B_{dyn}^2(x)}}=0
\label{25}\ed
together with the boundary conditions ($x=p^2$)
\bg
\left. {x{{dB_{dyn}(x)} \over {dx}}+B_{dyn}(x)} \right|_{x=\Delta
}=0;\quad\quad\left. {x^2{{dB_{dyn}(x)} \over {dx}}} \right|_{x=\delta }=0
\label{26}\ed
where $\Delta\rightarrow \inf$ and $\delta \rightarrow 0$ and
$\lambda =3\bar{g}^2C_2(N)/16\pi ^2$. We are interested in finding not so much
in existence
of non-trivial solution to (\ref{25}) but in determining when the trivial
solution
$B_{dyn}(x)=0$ is in fact a stable solution. To examine the stability of the
trivial
solution, let us linearize (\ref{25}) about $B_{dyn}=\varepsilon (x)$.  We find
\bg
x^2\varepsilon (x)+2x\varepsilon (x)+\lambda \kern 1ptx\varepsilon (x)=0
\label{27a}\ed
with
\bg
\left. {x\varepsilon (x)+\varepsilon (x)} \right|_{x=\Delta
}=0;\quad\quad\left. {x^2\varepsilon (x)} \right|_{x=\delta }=0
\label{27b}\ed
whose general solution is
\bg
\varepsilon (x) =c_1x^{\r _+}+c_2x^{\r _-} \label{28}
\ed
where
\bg
\rho _\pm =-{1 \over 2}\pm {1 \over 2}\sqrt {1-4\lambda }.
\label{29}\ed
$c_1$ and $c_2$ are to be determined by the boundary conditions. If $c_1$ and
$c_2$ are non-zero, then the solution $\varepsilon (x)\rightarrow \inf$ as
$x\rightarrow 0$
and thus the trivial solution is not stable. Thus stability requires
$c_1=c_2=0$. Substituting (\ref{28}) into the b.c.\ (\ref{27b}), we find the
following
condition for the trivial solution to be stable:
\bg
{{1+\sqrt {1-4\lambda }} \over {1-\sqrt {1-4\lambda }}}=({\delta  \over \Delta
})^{{{\sqrt {1-4\lambda }} \over 2}}.
\label{30}\ed
If $\lambda \leq 1/4$, a solution to (\ref{30}) is not possible since the left
hand side
of (\ref{30}) is finite while the right hand side tends to zero in the limit
$\delta \rightarrow 0$ and $\Delta \rightarrow \inf$. If $\lambda > 0$,
$\sqrt{1-4\lambda}$ is purely imaginary, then (\ref{30}) is a transcendental
equation  for $\lambda$ \cite{stam}. For fixed $\Delta$ and $\delta$, it has an
infinite set of solutions for $\lambda$ which becomes dense over the whole
domain
$\lambda >1/4$ as $\Delta /\delta$ becomes large. Hence the critical point is
$\lambda _c=1/4$ or $\alpha _c=\pi /4$. For $\lambda \leq \lambda _c$ we have a
stable
trivial solution $B_{dyn}(x)=0$. The critical current mass is
\bg
m_c=\Lambda _{QCD} \exp \frac{2\pi}{\beta _0\alpha _c}\simeq e\Lambda _{QCD}.
\label{31}
\ed
Implicit in the analysis above is the assumption  that there are no other
stable
solutions.

The solution to (\ref{24}) for $B_{kin}$ can be derived by the iteration
process while
$B_{dyn}=0$ must be substituted into (\ref{24}). To the first order one has
$B_{kin}=1+O(\bar{g}^2(t))$ and the kinematical part of total self-mass reads
from (\ref{21})
\bg
B_K^R(p;g(\mu ),m(\mu ),\xi (\mu );\mu )\cong m(\mu )(b\ln {m \over \mu
})^{-c/b}(1+O(\bar g^2))\quad\quad(m_P \gg \Lambda _{QCD})
\label{32} \ed
where $c=3C_2(N)/8\pi ^2$ and $b=\beta _0/8\pi ^2$.

We have concerned ourselves with the self-mass for the heavy quark. Having
assumed
the chiral phase transition for light quarks, we may draw some conclusions for
the self-mass
of the massive quark as the function of its current mass. In the range of
$m\simeq 0$, the $B_D^R$ dominates. As $m$ becomes large, the kinematical part
grows almost linearly and the dynamical part changes slowly. At $m_P=m_c$, the
dynamical part drops to zero and the total self-mass will exhibit a local
minimum. As $m$ gets even larger, the self-mass is completely governed by the
kinematical part (except for the possible contributions from the gluon
condensate)
and eventually blows up as shown in (\ref{32}) when $m\rightarrow \inf$. The
quark
condensate defined as
\bg
\left\langle {\bar qq} \right\rangle & = & -\int {{{d^4k} \over {(2\pi
)^4}}TrS_F(k)}\nonumber \\
  & = & -\int {{{d^4k} \over {(2\pi )^4}}Tr{{B_K^R} \over
{k^2-(B_K^R+B_D^R)^2}}}-\int {{{d^4k} \over {(2\pi )^4}}Tr{{B_D^R} \over
{k^2-(B_K^R+B_D^R)^2}}}\nonumber \\
  & = & \left\langle {\bar qq} \right\rangle_D+\left\langle {\bar qq}
\right\rangle_K
\label{34} \ed
is expected to observe the same drop when $B_D^R=0$.
  \pagebreak


\begin{thebibliography}{99}
\bibitem[1]{review}For reviews on this issue, see for example, M.E.\ Peskin,
SLAC-PUB-3021 (1982); V.A.\ Miransky and P.I.\ Fomin, Sov. J. Part. Nucl. 16,
203 (1985);
A. Barducci, R. Casalbouoni, S.D.\ Dominici and R.\ Gatto, Phys. Rev. D38, 238
(1988);
K.\ Lane, Phys. Rev. D10, 1353 (1974).
\bibitem[2]{pagels}H.\ Pagels,Phys. Rev. D19 (1979).
\bibitem[3]{lar}K.\ Langfeld, R. Alkofer and H.\ Reinhardt, Phys. Lett. B277,
163 (1992)
\bibitem[4]{zh}Z.\ Huang and K.S. Viswanathan, XIV Intl. Warsaw meeting on
Elem.
Particles: Puzzles on Electroweak Scale, Warsaw May 27-31 (1991), (World
Scientific).
\bibitem[5]{hvw}Z.\ Huang, K.S.\ Viswanathan and D.D.\ Wu, Mod. Phys. Lett. A6,
711 (1991).
\bibitem[6]{thooft} S.\ Weinberg, Phys. Rev. D8, 3497 (1973); G.\ t' Hooft,
Nucl. Phys. B61, 455 (1973).
\bibitem[7]{jbw}K.\ Johnson, M.\ Baker and R.\ Willey, Phys. Rev. B1111, 136
(1964); 1699, 163 (1967).
\bibitem[8]{narrison}S.\ Narison, Phys. Rep. C82, 263 (1982).
\bibitem[9]{fukuda}R.\ Fukuda and T.\ Kugo, Nucl. Phys. B117, 250 (1976).
\bibitem[10]{stam}D.\ Atikinson, P.W.\ Johnson and K.\ Stam, Phys. Lett. B201,
105 (1988).
\end{thebibliography}
\end{document}